\documentclass[12pt]{article}

\usepackage[english]{babel}
\usepackage{amssymb,amsmath,amsthm,graphicx}
\usepackage[round]{natbib}

\newtheorem{theorem}{Theorem}

\newtheorem{corollary}{Corollary}

\newtheorem{definition}{Definition}

\newtheorem{remark}{Remark}

\newtheorem{example}{Example}

\usepackage{geometry} 
\usepackage{graphicx} 

\usepackage{booktabs} 
\usepackage{array} 
\usepackage{paralist} 
\usepackage{verbatim} 
\usepackage{subfig} 
\usepackage{bbding}

\usepackage{fancyhdr} 
\usepackage{sectsty}
\allsectionsfont{\sffamily\mdseries\upshape} 

\usepackage[nottoc,notlof,notlot]{tocbibind} 
\usepackage[titles,subfigure]{tocloft} 
\newcommand{\real}[0]{\mathbb R}

\date{\today}

\title{Conditions for the uniqueness of the Gately point for cooperative games}
\author{Jochen Staudacher \Envelope , Johannes Anwander \\ Fakult\"at Informatik \\ Hochschule Kempten \\ 87435 Kempten, Germany \\ 
        Email (Corresponding Author): Jochen.Staudacher@hs-kempten.de}

\begin{document}
\maketitle
\begin{abstract} 
We are studying the Gately point, an established solution concept for cooperative games.
We point out that there are superadditive games 
for which the Gately point is not unique, i.e.~in general the concept is rather 
set-valued than an actual point.
We derive conditions under which the Gately point is guaranteed to be a unique imputation and provide a geometric interpretation. 
The Gately point can be understood as the intersection of a line defined by two 
points with the set of imputations. Our uniqueness conditions guarantee that these 
two points do not coincide. We provide demonstrative interpretations for negative propensities to disrupt. 
We briefly show that our uniqueness conditions for the Gately point include  
quasibalanced games and discuss the relation of the Gately point to the $\tau$-value 
in this context. Finally, we point out 
relations to cost games and the ACA method and end 
upon a few remarks on the implementation of the Gately point and an upcoming software package 
for cooperative game theory.

\emph{JEL-classification:} C71
\end{abstract}
\section{Introduction}
\label{intro}
Dermot Gately introduced a new solution concept for cooperative games with 
transferable utility in \cite{Gately} based on 
minimizing the temptation to leave the grand coalition for individual players. 
In the original paper \cite{Gately} the problem of sharing the gains from a joint investment 
in an electric power grid in India between the participating regions is resolved 
with the help of the concept ``equal propensity to disrupt''.   
Since the publication of \cite{Gately}, the so-called Gately point has become a 
well-established solution concept taught in books by \cite{Straffin} and \cite{Narahari} and  
mentioned in highly regarded survey articles, like e.g.~\cite{SandlerTschirhart} and \cite{Young}. \\
As of 6 January 2019, 211 quotes 
of 
\cite{Gately} can be found 
on GoogleScholar. From its name Gately point one is tempted to assume that the solution concept 
in question was always unique.
In this paper we point out that this is not actually the case. We strive to answer the following question: 
Under which conditions is the Gately point a unique imputation? Along the way, we also discuss what negative propensities to 
disrupt tell us about a cooperative game. 
%

\section{Preliminary definitions}
\label{sec:2}
We are studying a transferable utility game (TU game) in characteristic function form consisting of the 
player set $N =\{ 1, \dots , n \}$ and the characteristic
function $v: 2^{N} \to \real$ with $v (\emptyset) = 0$. We are using the shorthand notations 
\begin{displaymath}
v_{i} = v(\{ i \}) \quad \textrm{for} \quad i=1, \dots , n,
\end{displaymath}
for the worths of the singleton coalitions. 
\begin{definition} (see \cite{BranzeiBook}, p.~20)
The so-called utopia payoff of player $i$ is given by 
\begin{displaymath}
M_{i} = v(N) - v(N \backslash \{ i \}) \quad \textrm{for} \quad i=1, \dots , n,
\end{displaymath}
i.e.~$M_{i}$ is the 
marginal contribution of player $i$ to the grand coalition. 
\end{definition}
\noindent
In this article we will only study games satisfying essentiality in the sense of \cite{Chakravarty}, p.~23. 
\begin{definition} (see \cite{Chakravarty}, p.~23)
We call a transferable utility game with player set $N =\{ 1, \dots , n \}$ and characteristic
function $v: 2^{N} \to \real$ essential if
\begin{equation} \label{imput}
\sum_{j=1}^{n} v_{j} < v(N).
\end{equation}
\end{definition}
\noindent
The imputation set of any essential TU game is guaranteed to consist of more 
than a single point. For a solution concept in cooperative game theory one would normally prefer the solution vector $x \in \real^{n}$ 
to be an imputation, i.e.~both individually rational $x_{i} \geq v_{i}$ for all 
$i=1, \dots, n$ and efficient $\sum_{j=1}^{n} x_{j} = v(N)$. 
For a formal defintion of the imputation set we refer to 
\cite{PelegSudhoelter}, p.~20, or 
\cite{Narahari}, p.~407. \\ 
Note that any cooperative game satisfying (\ref{imput}) is strategically equivalent 
to a $0$-$1$-normalized game, see \cite{MaschlerSolanZamir}, p.~670, or \cite{Chakravarty}, p.~24. 
\begin{definition} (see e.g.~\cite{PelegSudhoelter}, p.~10)
We call a transferable utility game with player set $N =\{ 1, \dots , n \}$ and characteristic
function $v: 2^{N} \to \real$ 
weakly superadditive if 
\begin{equation} \label{weaklySuper}
v (S \cup \{i \} ) \geq v(S) + v_{i} \quad \textrm{for all} \quad S \subseteq N \quad \textrm{and} \quad i \notin S .  
\end{equation}
\end{definition}
\noindent
Note that weak superadditivity (\ref{weaklySuper}) guarantees 
\begin{equation} \label{condition1}
v_{i} \leq M_{i} \quad \textrm{for} \quad i=1, \dots , n.
\end{equation}
For later convenience we repeat the following 
\begin{definition} (see e.g. \cite{Straffin}, p.~131, or \cite{Narahari}, p.~408)
We call a transferable utility game with player set $N =\{ 1, \dots , n \}$ and characteristic
function $v: 2^{N} \to \real$ 
superadditive if
\begin{equation} \label{superAdd}
v (S \cup T ) \geq v(S) + v(T) \quad \textrm{for all} \quad S, T \subseteq N \quad \textrm{with} \quad S \cap T = \emptyset .  
\end{equation}
\end{definition}
\noindent
Finally, we would like to introduce the following game property.
\begin{definition}
We call a transferable utility game with player set $N =\{ 1, \dots , n \}$ and characteristic
function $v: 2^{N} \to \real$ 
weakly constant-sum if
\begin{equation} \label{weakConstantSum}
v_{i} + v(N \backslash \{ i \}) = v(N) \quad \textrm{for all} \quad i=1, \dots , n.
\end{equation}
\end{definition}
\noindent
Note that weakly constant-sum games $v$ can equivalently be characterized by  
\begin{equation} \label{conditionFail}
v_{i} = M_{i} \quad \textrm{for} \quad i=1, \dots , n.
\end{equation}
\section{Nonuniqueness of the Gately point and uniqueness conditions}
\label{sec:3}
In this section we will introduce the Gately point as a solution concept for cooperative games 
along the lines of the article by 
\cite{LittlechildVaidya}. \\
\noindent
The following definition is central to understanding the Gately point as a solution concept for cooperative games.
\begin{definition} (see \cite{LittlechildVaidya}, p.~152)
For a given transferable utility game with player set $N =\{ 1, \dots , n \}$ and characteristic
function $v: 2^{N} \to \real$  the expression 
\begin{equation} \label{propensity}
d(i,x) = 
\frac{ v(N) - v(N \backslash \{ i \}) - x_{i}}{x_{i} - v_{i}} = \frac{M_{i} - x_{i}}{x_{i} - v_{i}}
\end{equation}
quantifies the 
propensity to disrupt of player $i$ 
for a payoff vector $x \in \real^{n}$ in the interior of the imputation set, i.e.~$\sum_{j=1}^{n} x_{j} = v(N)$ 
with  $x_{i} > v_{i}$ for all $i=1, \dots, n$.
\end{definition} 
\noindent
Expression (\ref{propensity}) quantifies the disruption caused if player $i$ breaks away from the grand coalition.
Within (\ref{propensity}) the denominator stands for the loss incurred by player $i$ for breaking 
away from the grand coalition, whereas the numerator stands for the joint loss of the rest of the 
players due to the breakup caused by player $i$. \\
The original approach in \cite{Gately} for three-person games was generalized to $n$-person games 
by \cite{LittlechildVaidya}, p.~152. The idea is simply to find an imputation $x \in \real^{n}$ 
with minimal propensity to disrupt. 
It can be shown that this minimal propensity to disrupt can be found 
by equating the propensity to disrupt over all players, i.e.
\begin{displaymath}
d(i,x) = d^{*} \quad \textrm{for} \quad i=1, \dots , n.
\end{displaymath}
\noindent
As pointed out by \cite{LittlechildVaidya}, p.~153, using (\ref{propensity}) one can easily 
find the following closed-form expression 
\begin{equation} \label{equalProp}
d^{*} = \frac{(n-1) v(N) - \sum_{j=1}^{n} v(N \backslash \{ j \})}{v(N) - \sum_{j=1}^{n} v_{j}} = \frac{\sum_{j=1}^{n} M_{j} - v(N)}{v(N) - \sum_{j=1}^{n} v_{j}}
\end{equation}
which also highlights the fact that the Gately point is a solution concept depending solely 
on the values of the coalitions of sizes $1$, $n-1$ and $n$. \\
Looking at (\ref{propensity}) one recognizes that for $d^{*}=-1$ we can not solve for the Gately point.
This case can indeed occur for games satisfying (\ref{imput}) and (\ref{weaklySuper}). 
We formalize these findings in 
\begin{theorem}
For an essential transferable utility game with player set $N =\{ 1, \dots , n \}$ and characteristic
function $v: 2^{N} \to \real$ the Gately point is well-defined unless the equal propensity 
to disrupt $d^{*}=-1$. We can find the Gately point 
as the unique imputation $x \in \real^{n}$ with the components
\begin{equation} \label{Gately}
x_{i}  = v_{i} + (v(N) - \sum_{j=1}^{n} v_{j}) \frac{M_{i} - v_{i}}{\sum_{j=1}^{n} M_{j} - \sum_{j=1}^{n} v_{j}}
\end{equation}
for $i=1, \dots ,n$, if one of the following two conditions holds: \\
a) For games satisfying (\ref{condition1}) 
there needs to hold 
\begin{equation} \label{condition2}
v_{i} < M_{i} \quad \textrm{for at least one} \quad i \in \{ 1, \dots , n \} , 
\end{equation}
i.e.~(\ref{condition1}) is satisfied with strict inequality for at least one $i \in \{ 1, \dots , n \}$. \\
b) We also obtain the Gately point $x \in \real^{n}$ as a unique imputation if 
\begin{equation} \label{condition3}
v_{i} \geq M_{i} \quad \textrm{for} \quad i=1, \dots , n, 
\end{equation}
as long as (\ref{condition3}) is satisfied with strict inequality for at least one $i \in \{ 1, \dots , n \}$, i.e.~as 
long as the game is not weakly constant-sum (\ref{conditionFail}).
\end{theorem}

Proof: As long as $d^{*} \neq -1$ the expression (\ref{Gately}) can be found using (\ref{equalProp}) by simple algebra. 
When $d^{*} \neq -1$ it is justified to set $x_{i}=v_{i}$ for those $i \in \{ 1, \dots , n \}$ with $v_{i}=M_{i}$.
Looking at the expression (\ref{Gately}), essentiality (\ref{imput}) implies that 
$x \in \real^{n}$ is an imputation if and only if 
\begin{displaymath}
\frac{M_{i} - v_{i}}{\sum_{j=1}^{n} M_{j} - \sum_{j=1}^{n} v_{j}} \geq 0 
\end{displaymath}
for all  $i \in \{ 1, \dots , n \}$. The latter condition is fulfilled for both games satisfying (\ref{condition1}) 
and games satisfying (\ref{condition3}) as long as these games are not weakly constant-sum (\ref{conditionFail}). 

\begin{remark}
The case $d^{*}<0$ can be interpretated as enthusiasm of each player not to be the one left out of the grand coalition.
In other words: $d^{*}<0$ indicates that coalitions of size $n-1$ are preferred over the grand coalition. 
In the case of (\ref{condition3}) being satisfied with strict inequality for at least one $i \in \{ 1, \dots , n \}$ 
this fact is particularly striking as there even holds $d^{*}<-1$.
\end{remark}
\begin{remark}
Geometrically, (\ref{Gately}) allows us to interpret the Gately point as the intersection of the 
imputation set with the half-line drawn from the point $(v_{1}, \dots , v_{n})$ 
with directional vector $(M_{1} - v_{1}, \dots , M_{n} - v_{n})$.
\end{remark}
\begin{remark}
For $0$-normalized games (\ref{Gately}) simplifies to 
\begin{equation} \label{Gately0}
x_{i}  = v(N) \frac{M_{i}}{\sum_{j=1}^{n} M_{j}} 
\end{equation}
for $i=1, \dots ,n$. 
\end{remark}

\noindent
We finally consider 
\begin{example} \label{Ex1}
Let the three-person game $v$ be given by 
\begin{displaymath}
v_{1}=3, v_{2}=4, v_{3}=5, v(\{ 1,2 \} )=9, v(\{ 1,3 \} )=10, v(\{ 2,3 \} )=11, v(N)=14.
\end{displaymath}
\end{example}
The above game is clearly superadditive (\ref{superAdd}) and essential (\ref{imput}), 
but the propensity to disrupt equals $-1$ for 
every imputation $x$. In a sense, the Gately point for $v$ would be the complete imputation set. 
Naturally, one would make the identical observation considering the 
$0$-normalization of $v$, i.e.~the coalitional game $w$ with 
$w_{1}=w_{2}=w_{3}=0, w(\{ 1,2 \} )=w(\{ 1,3 \} )=w(\{ 2,3 \} )=w(N)=2$, 
or the $0$-$1$-normalization of $v$, i.e.~the coalitional game $u$ with 
$u_{1}=u_{2}=u_{3}=0, u(\{ 1,2 \} )=u(\{ 1,3 \} )=u(\{ 2,3 \} )=u(N)=1$.
Note that the latter could also be interpreted as a weighted voting game.

\section{Relations to the $\tau$-value}
\label{sec:4}
In the previous section we have seen that the Gately point is the intersection of the 
imputation set with a line connecting the points $(v_{1}, \dots , v_{n})$ 
and $(M_{1}, \dots , M_{n})$ and pointed out a problem for the case that these two points 
coincide (\ref{conditionFail}). There is another well-established solution concept in cooperative game theory 
computing the intersection of a line connecting two points with the imputation set, i.e.~the 
$\tau$-value proposed by \cite{TijsTau}. 
\begin{definition} (see \cite{BranzeiBook}, p.~20)
The remainder $R(S, i)$
of player $i$ in coalition $S$ is the amount which remains for player $i$ if
coalition $S$ forms and the rest of the players in coalition $S$ all obtain their 
individual utopia payoffs, i.e.
\begin{displaymath}
R(S,i) = v(S) - \sum_{j \in S, j \neq i} M_{j}. 
\end{displaymath}
We can define a vector of minimal rights with components 
\begin{displaymath}
m_{i} = \max_{S: i \in S} R(S,i),  \quad \textrm{for} \quad i=1, \dots , n,
\end{displaymath}
since player $i$ has a justification to ask at least $m_{i}$ in the grand coalition. 
\end{definition}
\noindent
The $\tau$-value is defined only for quasibalanced games. 
\begin{definition} (see e.g.~\cite{BranzeiBook}, pp.~31)
We call a transferable utility game with player set $N =\{ 1, \dots , n \}$ and characteristic
function $v: 2^{N} \to \real$ quasibalanced if 
\begin{equation} \label{quasi1}
m_{i} \leq M_{i} \quad \textrm{for all} \quad i \in \{ 1, \dots , n \} 
\end{equation}
and 
\begin{equation} \label{quasi2}
\sum_{j=1}^{n} m_{j} \leq v(N) \leq \sum_{j=1}^{n} M_{j}.  
\end{equation}
For a quasibalanced game $v$ the $\tau$-value is defined as the 
intersection of the 
imputation set with the line from the minimal rights vector  
$(m_{1}, \dots , m_{n})$ to the  
utopia payoff vector $(M_{1}, \dots , M_{n})$. 
\end{definition}
\begin{remark} (see e.g.~\cite{BranzeiBook}, p.~32)
We can find  
the $\tau$-value with the components 
\begin{displaymath}
\tau_{i}  = \alpha m_{i} + (1- \alpha) M_{i}, 
\end{displaymath}
where $\alpha \in [0,1]$ is uniquely determined by the condition $\sum_{i=1}^{n} \tau_{i} = v(N)$. 
\end{remark}
\noindent
Combining (\ref{equalProp}) and condition (\ref{quasi2}) we find that 
for quasibalanced games $d^{*} \geq 0$ is guaranteed and we 
arrive at
\begin{corollary}
The Gately point is always unique for quasibalanced games.
\end{corollary}
\noindent
Note that the conditions we formulated for 
the Gately point to be a unique imputation are more general than quasibalancedness, i.e.~there 
are games for which the $\tau$-value is not defined whereas the Gately point is. Consider 
\begin{example} 
Let the three-person game $v$ be given by 
\begin{displaymath}
v_{1}=3, v_{2}=4, v_{3}=5, v(\{ 1,2 \} )=9, v(\{ 1,3 \} )=10, v(\{ 2,3 \} )=11, v(N)=14.5.
\end{displaymath}
\end{example}
The game $v$ is not quasibalanced and its 
Gately point can be computed to $x_{1}= 3 \frac{5}{6}, x_{2}= 4 \frac{5}{6}, x_{3}= 5 \frac{5}{6}$. \\
We finally observe that the problem we report for the Gately point never occurs for the 
$\tau$-value which we already mentioned to be the intersection of the 
imputation set with a line drawn from the point $(m_{1}, \dots , m_{n})$ 
to the point $(M_{1}, \dots , M_{n})$, see \cite{TijsTau}. However, if these two points 
coincide, then (\ref{quasi2}) guarantees this point to be an imputation and thus the 
$\tau$-value of the game. Note that in this special case there is  $d^{*} = 0$.

\section{Application to cost games and relations to the ACA-method}
\label{sec:5}
We are looking at cost games in characteristic function form consisting of the 
set $N =\{ 1, \dots , n \}$ of agents (or purposes, projects or services) and the characteristic
function $c: 2^{N} \to \real$ with $c (\emptyset) = 0$. We are using the shorthand notation 
\begin{displaymath}
c_{i} = c(\{ i \}) \quad \textrm{for} \quad i=1, \dots , n,
\end{displaymath}
for the costs of single agents. The connection to TU games is given by the 
associated savings game $v$ for $N =\{ 1, \dots , n \}$ defined by 
\begin{displaymath}
v(S) = \sum_{i \in S} c_{i} - c(S)
\end{displaymath}
for every coalition $S$. Note that the associated savings game $v$ is automatically 
$0$-normalized. \\
We are now discussing the so-called ACA (Alternate Cost Avoided) method, 
i.e.~an established method for cost allocation going back to \cite{TVA},   
along the lines of \cite{StraffinHeaney}. The ACA method has been 
widely discussed, see also \cite{Otten}, \cite{TijsDriessen} and \cite{Young}. \\
The ACA method is based on the concept of allocating separable costs 
\begin{displaymath}
SC_{i} = c(N) - c(N \backslash \{ i \}) = c_{i} - M_{i} 
\end{displaymath}
for each agent $i=1, \dots , n$. The remaining nonseparable costs 
\begin{equation} \label{NSC}
NSC = c(N) - \sum_{j=1}^{n} SC_{j} = \sum_{j=1}^{n} M_{j} - v(N) 
\end{equation}
are assigned in proportion to $c_{i} - SC_{i}$, i.e.~the final cost allocation 
for an individual agent $i$ is 
\begin{displaymath}
y_{i} = SC_{i} + \frac{c_{i} - SC_{i}}{\sum_{j=1}^{n} c_{j} - SC_{j}} NSC.
\end{displaymath}
As pointed out in \cite{StraffinHeaney}, p.~40, the corresponding savings 
allocation $x \in \real^{n}$ is exactly the Gately point, i.e.
\begin{displaymath}
x_{i} = c_{i} - y_{i} =  v(N) \frac{M_{i}}{\sum_{j=1}^{n} M_{j}} 
\end{displaymath}
as seen in (\ref{Gately0}). \\
It is very natural to understand why the problem of nonuniqueness of ACA never 
came up in the context of cost games. In practice, only subadditive cost 
games are studied, i.e.~the corresponding savings game is superadditive (\ref{superAdd}), 
see \cite{Young}, p.~1197. 
The ACA method can only fail to deliver 
a unique cost allocation if $M_{i}=0$, or equivalently 
$c_{i} = SC_{i}$, for $i=1, \dots , n$. Then (\ref{imput}) 
implies $NSC < 0$, whereas studies of ACA for good reason assume   
nonnegativity of nonseparable costs, see \cite{Otten}, p.~177, 
and \cite{TijsDriessen}, p.~1019. Practical ACA calculations would normally stop 
if $NSC < 0$ and this implies $d^{*} <0$, see (\ref{equalProp}) and (\ref{NSC}). 
We finally consider
\begin{example} \label{Ex3}
Let the subadditive three-agent cost game $c$ be given by 
\begin{displaymath}
c_{1}=7, c_{2}=8, c_{3}=9, c(\{ 1,2 \} )=14, c(\{ 1,3 \} )=15, c(\{ 2,3 \} )=16, c(N)=23.
\end{displaymath}
\end{example}

The corresponding savings game $u$ is the weighted voting game 
$u_{1}=u_{2}=u_{3}=0, u(\{ 1,2 \} )=u(\{ 1,3 \} )=u(\{ 2,3 \} )=u(N)=1$
we already know from Example \ref{Ex1}. The Gately point does not exist and so ACA fails 
to deliver a unique cost allocation. \\

\noindent
In general, ACA can only run into problems if all coalitions of size $n-1$ 
and the grand coalition make identical savings. Then we would expect a coalition of 
size $n-1$ to form, but we can not use ACA to single out the one agent $i$ to be 
left out. 

\section{Final remarks}
\label{sec:6}
The main purpose of this article is to answer the question when it is at all sensible to 
compute the Gately point of a TU game $v$. We derived very general conditions 
for the Gately point to be a unique imputation and pointed out why weakly constant-sum 
games lead to problems. We feel that our analysis underlines 
the criticism of the Gately point made in \cite{LittlechildVaidya}, p.~153, that 
the solution concept only makes use of the values of the coalitions of sizes $1$, $n-1$ and $n$ 
and completely ignores the rest of the information contained in the coalition function $v$. \\
The nonuniqueness of the Gately point was first discussed in \cite{Anwander} and 
it was discovered during efforts to implement the Gately point in R. 
The authors are currently finalizing an R-package named CoopGame (see \cite{CoopGame}) which the authors  
hope to make publicly available via CRAN, the Comprehensive R Archive Network.
Among various other solution concepts, the package CoopGame will not only 
provide an implementation of the Gately point, but also provide the user with 
possibilities to compute the equal propensity to disrupt $d^{*}$ of a given 
cooperative game $v$.
The scope of our Gately point implementation is slightly broader as  
for an inessential game $v$ in the sense of \cite{Narahari}, p.~408, i.e.~if (\ref{superAdd}) 
holds with  $\sum_{j=1}^{n} v_{j} = v(N)$, 
our code will simply return $(v_{1}, \dots , v_{n})$. Otherwise, we make 
sure to check the conditions derived in this paper before 
the computation of the Gately point and to return  
a meaningful message in the case the user specifies a 
TU game $v$ with an equal propensity to disrupt $d^{*}=-1$.

%
%
%


\bibliographystyle{spbasic}      


\end{document}